\documentclass{iaus}
\usepackage{graphicx}

\title[UCDs -- more massive than allowed?] 
{Ultra-compact dwarf galaxies -- more massive than allowed?}

\author[M. Hilker et al.]   
{Michael Hilker$^1$,
 S. Mieske$^1$, H. Baumgardt$^2$ \and J. Dabringhausen$^2$}

\affiliation{$^1$ESO, Karl-Schwarzschild-Str.\,2, 85748 Garching bei M\"unchen,
Germany \break email: mhilker@eso.org, smieske@eso.org\\[\affilskip]
$^2$AIfA, Universit\"at Bonn, Auf dem H\"ugel
71, 53121 Bonn, Germany \break email: holger@astro.uni-bonn.de, 
joedab@astro.uni-bonn.de}

\pubyear{2007}
\volume{246}  
\pagerange{119--126}
\date{?? and in revised form ??}
\jname{Dynamical Evolution of Dense Stellar Systems}
\editors{E. Vesperini, M. Giersz \& A. Sills, eds.}
\begin{document}

\maketitle

\begin{abstract}
Dynamical mass estimates of ultra-compact dwarfs galaxies and massive
globular clusters in the Fornax and Virgo clusters and around the giant
elliptical Cen\,A have revealed some surprising results: 1) above $\sim10^6 
M_\odot$ the mass-to-light ($M/L$) ratio increases with the objects' mass; 
2) some UCDs/massive GCs show high $M/L$ values (4 to 6) that are not 
compatible with standard stellar population models; and 3) in the 
luminosity-velocity dispersion diagram, UCDs deviate from the well defined 
relation of ``normal'' GCs, being more in line with the Faber-Jackson    
relation of early-type galaxies.
In this contribution, we present the observational evidences for high 
mass-to-light ratios of UCDs and discuss possible explanations for them.
\end{abstract}

\firstsection 
\section{Introduction}

The so-called ultra-compact dwarf galaxies (UCDs) are very massive 
($10^6 M_\odot<M<10^8 M_\odot$), old, compact stellar systems that were 
discovered in nearby galaxy clusters about a decade ago (\cite{hilk99}, 
\cite{drin00}). Their nature is unknown yet. Maybe they are remnant 
nuclei of disrupted galaxies, or maybe they are merged stellar super-clusters
formed in interacting galaxies. Regardless of what UCDs actually are, some
properties divide them from ``ordinary'' globular clusters (GCs). The 
half-light radii of UCDs scale with luminosity reaching $\sim90$ pc for the 
most massive UCDs. Unlike for GCs, their densities within the half-light 
radii are not increasing with mass but stay at a constant level or even 
decrease. Thus UCDs are not that compact at all when compared to $10^6 
M_\odot$ GCs, but certainly much denser than dwarf ellipticals of
comparable mass.

\section{Mass determinations and results}

To estimate the masses of UCDs a new modelling program has been developed 
that allows a choice of different representations of the surface brightness 
profile (i.e. Nuker, Sersic or King laws) and corrects the observed velocity 
dispersions for observational parameters (i.e. seeing, slit size). The 
derived dynamical masses are compared to those expected from stellar 
population models. For more details, see \cite{hilk07}.

The masses, central densities and mass-to-light ($M/L$) ratios of different 
hot stellar systems (GCs, UCDs, dEs, bulges and ellipticals) were compared
with each other (Dabringhausen et al. 2007, in prep.). The findings are as
follows: 1) In the central density vs. mass plane, there seems to be an 
upper limit of about $10^4 M_\odot /pc^3$ for GCs of $\sim 10^6 M_\odot$.
UCDs scatter towards lower densities with increasing mass. 2) In the $M/L$ vs.
mass plane, the $M/L$ ratio icreases with mass above $\sim 10^6 M_\odot$
(see Fig.\,1, right panel),
reaching values typical for bulges and ellipticals. 3) When plotting a
normalised $M/L$ ratio (taking out the metallicity dependence on $M/L$) vs.
mass, the objects more massive than a few times $10^6 M_\odot$ show 
systematically higher $M/L$ values than the lower mass `normal' GCs. These
high values cannot easily be explained with standard single stellar population
models (see Fig.\,1, right panel).
Interestingly, the transition from low-$M/L$ to high-$M/L$ objects corresponds
to the mass regime ($10^6$-$10^7 M_\odot$) where the relaxation time at the
half-light radius exceeds a Hubble-time.

\begin{figure}
\centering
\resizebox{6.5cm}{!}{\includegraphics{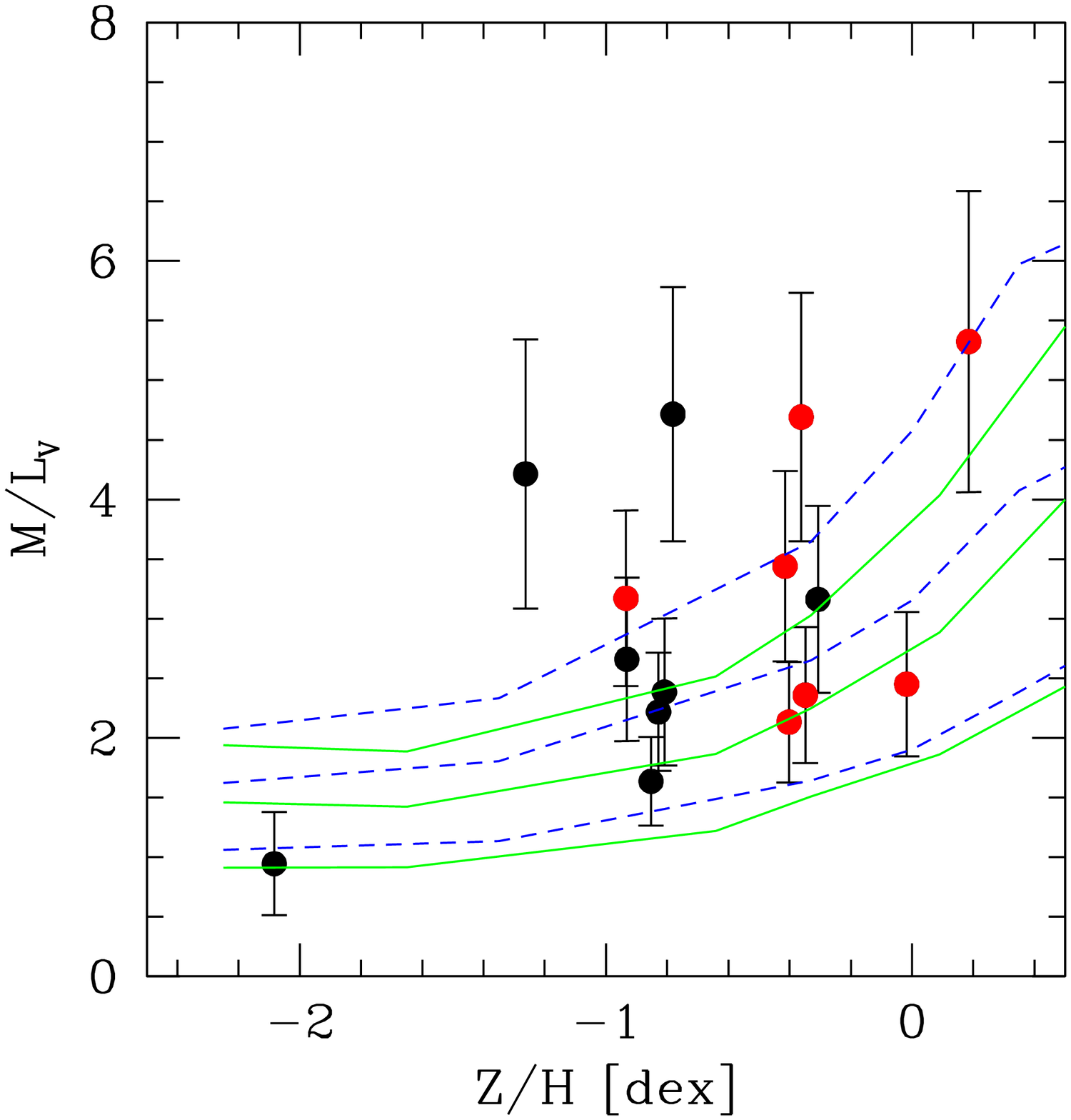} }
\resizebox{6.5cm}{!}{\includegraphics{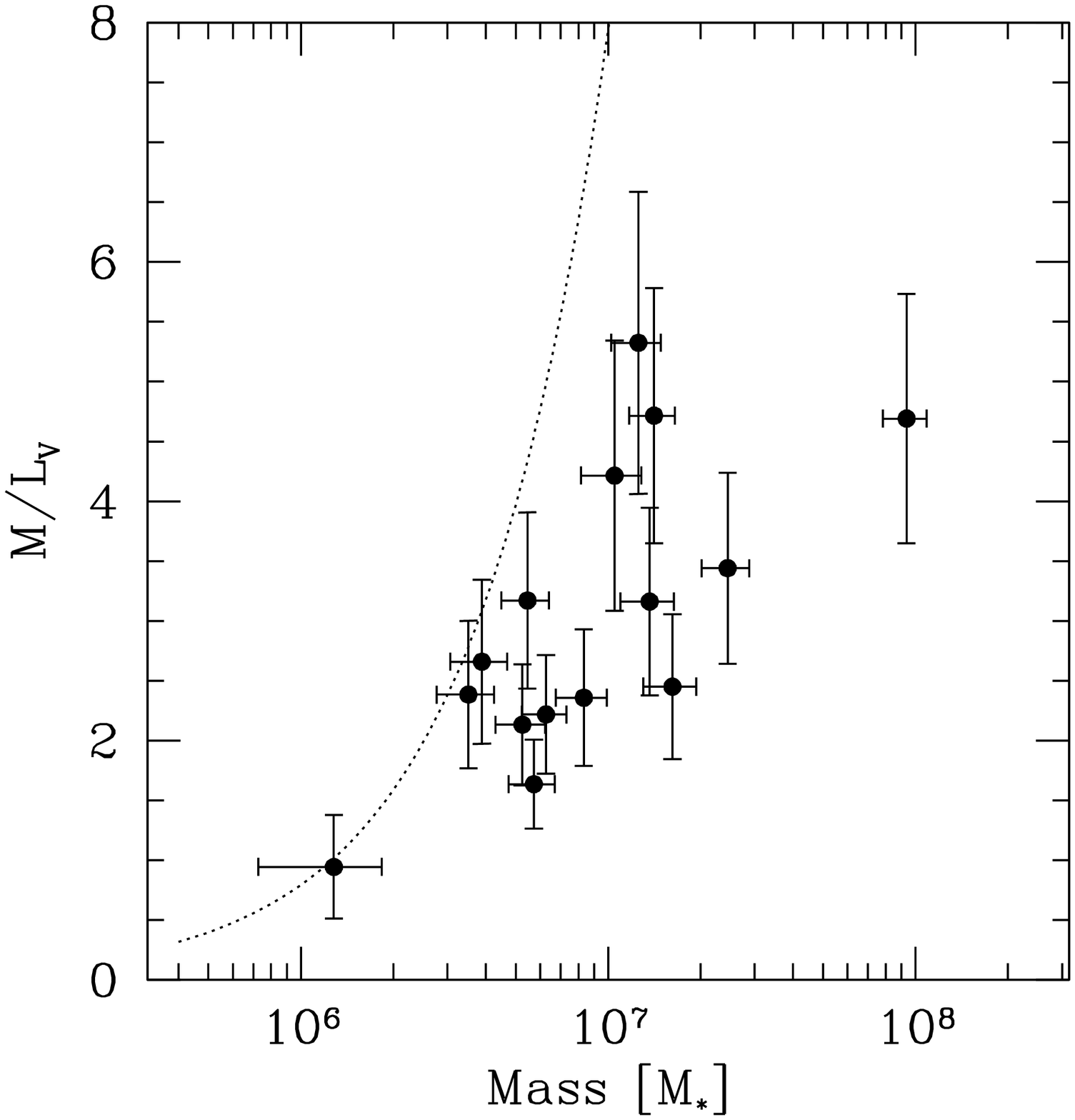} }
\caption[]{Most recent $M/L_V$ determinations for Fornax UCDs from FLAMES/UVES
observations (Mieske et al., 2008). Left: SSP models (5, 9 and 13 Gyr) 
from Bruzual \& Charlot (2003, solid) and Maraston (2005, dashed).
Right: The dotted line indicates the observational limit.}
\end{figure}

\section{Possible explanations for unusually high M/L ratios}

\vskip1mm
\noindent
{\bf 1) Dark matter:}
This would imply a very high DM densitiy within the core radius of UCDs.
A cuspy NFW halo with $10^8$-$10^{12} M_\odot$ would be needed. Might UCDs be 
surviving dense low mass DM sub-structures?
 
\vskip1mm
\noindent
{\bf 2) Tidal heating:}
UCDs might be out of dynamical equilibrium (\cite{fell06}). 
However, very eccentric orbits would be needed to observe high $M/L$-UCDs.

\vskip1mm
\noindent
{\bf 3) Top-heavy IMF:}
Remnants of massive stars (stellar BHs, neutron stars and white dwarfs) might
contribute to the unseen mass and increase the $M/L$ value. An IMF slope of 
$\alpha = -1$ to $-1.5$ for $M>1M_\odot$ would be needed.

\vskip1mm
\noindent
{\bf 4) Bottom-heavy IMF:}
Many low mass stars might contribute to the high $M/L$ value ($\alpha = 
-2.35$ low mass slope might explain it).

\vskip1mm
For a detailed discussion of these points, see Dabringhausen et al. (2007,
in prep.) and Mieske et al. (2007, in prep.).

\end{document}